\documentclass[aps,pre,floats,floatfix,twocolumn,fleqn]{revtex4-2}

\usepackage{graphicx}
\usepackage{epstopdf}
\graphicspath{{./}{./Figs/}} 

\usepackage{braket}
\usepackage{latexsym}
\usepackage{amsmath}
\usepackage{amssymb}
\usepackage{bm}
\usepackage{wasysym}

\usepackage{ifthen}
\usepackage{color}
\usepackage[colorlinks,allcolors=blue,bookmarks=true]{hyperref}

\newcommand{\trc}{\mbox{trace}}

\newcommand{\amatrix}[1]{\begin{matrix} #1 \end{matrix}}

\renewcommand{\bra}[1]{\left\langle #1 \right|}
\renewcommand{\ket}[1]{\left| #1 \right\rangle}
\renewcommand{\braket}[1]{\left\langle #1 \right\rangle }

\renewcommand{\Braket}[2]{\left\langle #1 \middle| #2 \right\rangle}
\newcommand{\BraKet}[3]{\left\langle #1 \middle| #2 \middle| #3 \right\rangle}

\newcommand{\beq}{\begin{eqnarray}}
\newcommand{\eeq}{\end{eqnarray}}

\newcommand{\hide}[1]{}  

\newcommand{\Eq}[1]{{\textcolor{blue}{Eq.}}~\!\!(\ref{#1})} 
\newcommand{\Sec}[1]{{\textcolor{blue}{Sec.}}~(\ref{#1})} 
 
\newcommand{\Fig}[1] {{\textcolor{blue}{Fig.}}~\!\!\ref{#1}}
\newcommand{\sect}[1]{{\bf #1.-- }}

\newcommand{\hrefl}[2]{\href{#2}{(#1)}}

\begin{document}
\title{Quantum thermalization and the route to ergodicity} 

\author{Amichay Vardi$^{1,2}$} 
\author{Doron Cohen$^3$}
\affiliation{$^1$Department of Chemistry, Ben-Gurion University of the Negev, Beer-Sheva 84105, Israel}
\affiliation{$^2$ITAMP, Harvard-Smithsonian Center for Astrophysics, Cambridge, MA 02138, USA}
\affiliation{$^3$Department of Physics, Ben-Gurion University of the Negev, Beer-Sheva 84105, Israel 
}

\begin{abstract}
We consider a minimal model for quantum thermalization of coupled chaotic subsystems. The route towards ergodicity is explored as a function of the coupling strength. The results are contrasted with the predictions of standard Random Matrix Theory (RMT) and the Eigenstates Thermalization Hypothesis (ETH). We highlight a coupling regime of disparity between the spectral statistics that indicates chaos, and ergodicity measures that indicate lack of ETH thermalization. The analysis involves a revision of the energy shell concept, in a way that is consistent but independent of the semiclassical perspective.   
\end{abstract}

\maketitle

\section{Introduction}

Random matrix theory (RMT), and specifically the theory of Wigner banded random matrices (WBRM) \cite{wigner,wbr}, have been proposed as a framework to describe ergodization and thermalization. Consider the quantum thermalization between coupled subsystems of a bipartite disordered system, where each subsystem by itself is generically chaotic. The Hamiltonian is ${H = H_0 + \lambda V}$ where the inter-system interaction $\lambda V$ mixes the eigenstates of the uncoupled subsystems.

It has been previously realized  that generic perturbations of a quantized chaotic system (QCS) can be modeled by WBRM \cite{mario1,mario2,mario3,mario4,flamb,felix1,felix2}. It follows that for relatively weak coupling the mixing of the levels is described by the Wigner Lorentzian, while for strong coupling the line-shape becomes non-perturbative. 
Namely, for WBRM it is a semicircle line-shape, while for a QCS it is a semiclassical line-shape that reflects the underlying phase-space geometry. The latter is implied by the association of chaotic eigenstates with micro-canonical distributions in Berry's conjecture and the eigenstate thermalization hypotheis (ETH) \cite{Berry77,Deutsch91,Srednicki94,felix3,wls,lds,Olshanii}.       

Later studies have considered the route to ergodicity and thermalization of large arrays, with special emphasis on 1D chains \cite{Chain1,Chain2,Chain3,Chain4}, utilizing both spectral and novel ETH measures \cite{ETH1,ETH2}. In such studies $H_0$ is integrable while  ${\lambda V}$ spoils the integrability. Here, we consider a more appropriate modeling of thermalization, closer in spirit to the traditional thermodynamic concept, where Boltzmann's `molecular chaos' applies equally well to the constituent thermalizing subsystems. Accordingly, we consider below a  model for thermalization, where ${\lambda V}$ is the coupling between the quantized {\em chaotic} subsystems. Each subsystem is chaotic by itself, while the coupling allows exchange of particles and energy. This configuration may be thought of as `vintage thermalization' because it reflects the traditional 19th century reasoning of thermodynamics. Similar agenda has been presented in \cite{Tomsovic}.

\begin{figure}[t!]
\includegraphics[width=\columnwidth]{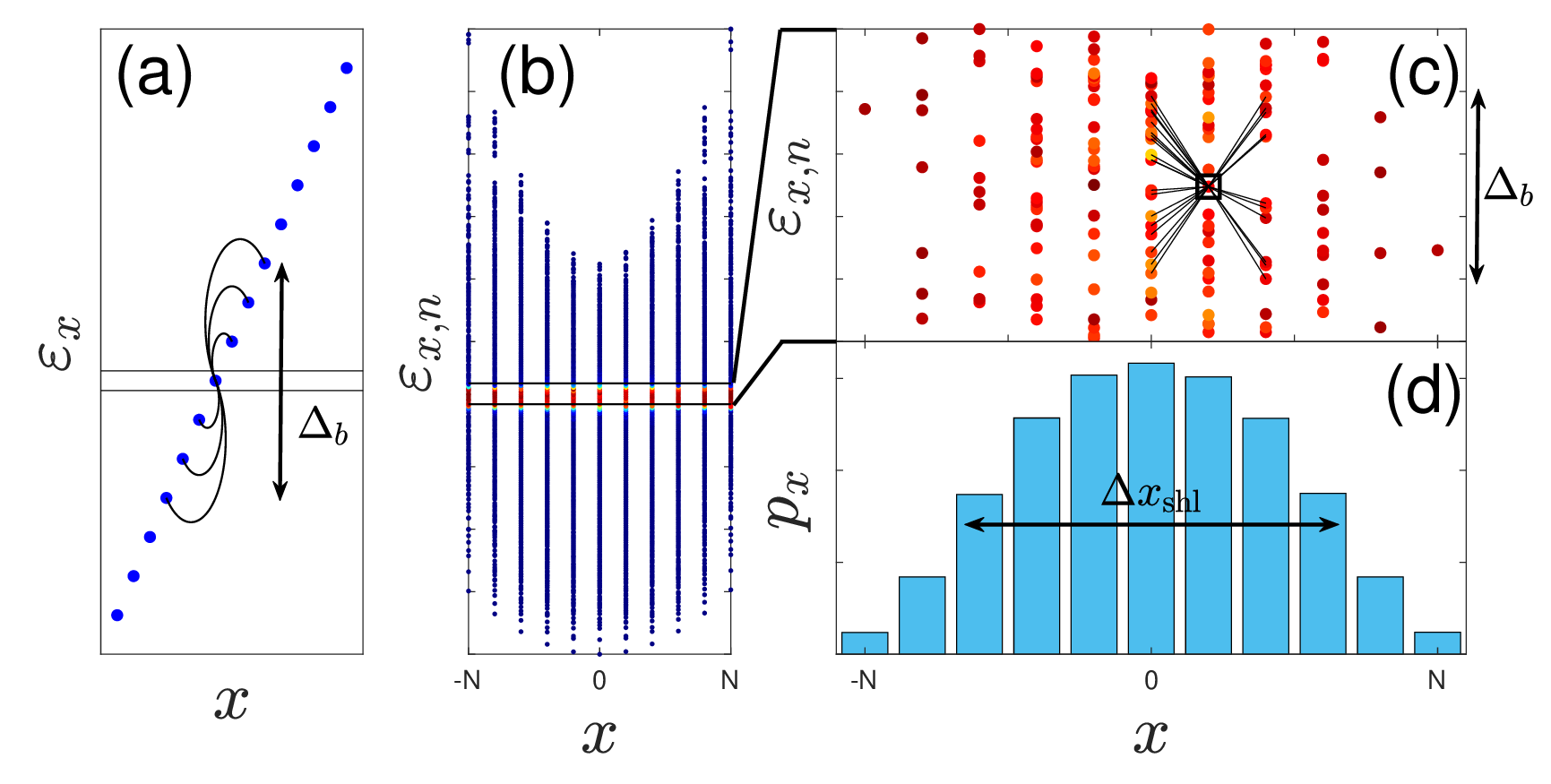}
\caption{
{\bf The unperturbed basis.}
The unperturbed energies $\varepsilon_{x}$ and $\varepsilon_{x,n}$ for the WBRM (a) and for the BHM (b,c), respectively. The bandwidth $\Delta_b$ is the maximal range in energy that can be reached by a single transition from an initial state. Possible first-order transitions are indicated.
The width of the energy-shell $\Delta x_{\text{shl}}$ is defined via the microcanonical distribution $p_x$ (d). It is not sensitive to~$\lambda$ for the BHM, and has a well defined finite value in the limit of zero coupling. In contrast, the WBRM model can be regarded as a singular limit that features  ${\Delta x_{\text{shl}} \propto \lambda}$.        
}  
\label{fg1}  
\end{figure}

As shown below, the application of WBRM theory to vintage thermalization is rather subtle. For illustration we consider a bi-partite Bose-Hubbard model (BHM) \cite{Gersch63,Jaksch98,Jaksch05,Essler05,Ponomarev11,trm,mlc,bhe}. The eigenstates of the disconnected subsystems $H_0|x,n\rangle=\varepsilon_{x,n}|x,n\rangle$ are direct products of subsystem eigenstates with well-defined integer subsystem occupations, indexed by two good quantum numbers: the occupation difference $x$ and the index $n$ of the state within the factorized spectrum (for a given $x$). These are plotted in \Fig{fg1}, where each point represents an unperturbed eigenstate.  The coupling $\BraKet{x,n}{V}{x',n'}$ connects only states with ${|x-x'|=1}$.  This 2D geometry of the basis for the BHM is very different from the 1D geometry of the WBRM (left panel of \Fig{fg1}), with substantial implications on the route to ergodicity. \\

\sect{Outline}
We introduce the model and basic definitions in Sections~II and~III. Critical discussion of the energy-shell concept is provided in Section~IV. Identification of chaos and ergodicity are explained in Sections~V and~VI. Then we explore the route towards ergodicity for the BHM in Section~VII, and contrast it with the WBRM. Auxiliary issues related to ergodicity measures and regime boundaries, are discussed in Sections~VIII and Section~IX. Finally, discussion and summary are provided in Section~X.   
Appendix~A provides a terse summary of the WBRM regimes, while Appendix~B presents extra numerical demonstration for the irrelevance of subsystem geometry.

\section{The BHM}

Consider the bipartite BHM for $N$ particles in $M=M_A{+}M_B$ sites, where $M_{\alpha}$ is the number of sites in subsystem $\alpha=A,B$. The Hamiltonian is,
\beq
H=\sum_{\alpha=A,B} H_\alpha+\lambda V
\eeq
where the subsystem Hamiltonians are,
\beq \nonumber
H_{\alpha}&=&\sum_i \varepsilon_{\alpha,i} \hat n_{\alpha,i}
+\frac{U}{2}\sum_{i} {\hat n}_{\alpha,i}({\hat n}_{\alpha,i}{-}1)
\\ 
~&~&-K\left(\sum_{\langle i,j\rangle} {\hat a}_{\alpha,i}^\dag {\hat a}_{\alpha,j}
+{\rm H.c.}\right)
\eeq
and the inter-system coupling is,
$$
V={\hat a}_{A,1}^\dag{\hat a}_{B,1} + {\hat a}_{A,M_A}^\dag{\hat a}_{B,M_B} +{\rm H.c.}
$$
In the above, the boson operator ${\hat a}_{\alpha,i}$ annihilates a particle in the $i$-th site of the $\alpha$ subsystem with population ${\hat n}_{\alpha,i}={\hat a}_{\alpha,i}^\dag{\hat a}_{\alpha,i}$, the intra-system hopping frequency is~$K$, and $\langle i,j\rangle$ denotes nearest-neighbor pairs. The coupling strength is~$U$, and the weak on-site random potential $\varepsilon_{\alpha,i}$ breaks parity.
In the numerical demonstrations we have ${M_A{=}M_B{=}4}$, $N{=}10$, and ${|\varepsilon_{\alpha,i}|<0.05}$, while ${K{=}1}$, and $U{=}1$. 

The Hilbert-space dimension of the BHM is,
\beq
D_M(N) = \frac{(N{+}M{-}1)!}{N!(M{-}1)!} = \sum_{x=-N}^N D_x 
\eeq
where $D_x=D_{M_A}\left(\frac{N{+}x}{2}\right)\times D_{M_B}\left(\frac{N{-}x}{2}\right)$ is the dimension of the $x$ sector of the spectrum (i.e. the number of states in each column of the diagram in \Fig{fg1}) and $x=-N,-N{+}2,\cdots,N{-}2,N$. In the absence of inter-system coupling ($\lambda=0$), the eigenenergies of the unperturbed Hamiltonian $H_0=H_A+H_B$ are $\varepsilon_{x,n}$, where the running index ${n=1,\dots,D_x}$ labels the $x$-sorted unperturbed energies.

\section{Eigenstates of the coupled subsystems}

Once the coupling $\lambda$ it turned on, the subsystems can exchange particles, and the coordinate $x$ is no longer a good quantum number. Accordingly we use a running index ${\nu=1,\cdots,D_M(N)}$ 
to label the many-body eigenstates and the respective eigenenergies $E_{\nu}$.
In the presentated results below, these energies are scaled as ${E:= (E{-}E_{\rm min})/(E_{\rm max}{-}E_{\rm min})}$, where $E_{\rm min}$ and $E_{\rm max}$ are the ground and top energies.

Each exact eigenstate features a probability distribution over the unperturbed states, namely, 
\beq
p_{x,n}^{(\nu)} \ \ = \ \ \Big| \Braket{x,n}{E_{\nu}} \Big|^2 
\label{xndist}
\eeq
The associated population-imbalance and energy distributions are,
\beq \label{epx}
p_x^{(\nu)} \ &=& \  \sum_n p_{x,n}^{(\nu)}
\label{xdist}
\\ \label{epe}
p^{(\nu)}(\varepsilon) \ &=& \ \sum_{x,n} p_{x,n}^{(\nu)} \, \delta(\varepsilon-\varepsilon_{x,n})
\label{edist}
\eeq
Using the above probability distributions,  
we define the following measures for the eigenstates, 
characterizing their spreading over the unperturbed basis: 
\beq
P_{\nu} &=& 
\max_{x,\nu} \Big\{ p_{x,n}^{(\nu)} \Big\} 
\ = \ \text{Peak probability}
\\
\mathcal{M}_{\nu} &=& 
\left[ \sum_{x,n} \big| p_{x,n}^{(\nu)} \big|^2 \right]^{-1} 
\!\! = \text{Participation number} \ \ \ \ \  
\\
\Delta x_{\nu} &=& \text{dispersion of $x$ via \Eq{epx}}
\\
\Delta {\varepsilon}_{\nu} &=& \text{dispersion of $\varepsilon$ via \Eq{epe}}
\eeq
The peak probability indicates whether the exact eigenstate $|E_{\nu}\rangle$ is a first-order perturbed eigenstate $|x,n\rangle$ of the uncoupled system. The participation number estimates the number of unperturbed eigenstates that contribute to the exact eigenstate. The dispersions quantify the spread in $x$ and in $\varepsilon$. It should be noted that while the full energy dispersion of an exact eigenstate is identically zero, $\Delta\varepsilon_\nu$  is the dispersion over the {\em unperturbed} energies, i.e. $\Delta\varepsilon_\nu=[\langle H_0^2\rangle_\nu-\langle H_0\rangle_\nu^2]^{1/2}$

\begin{figure}
\includegraphics[width=\columnwidth]{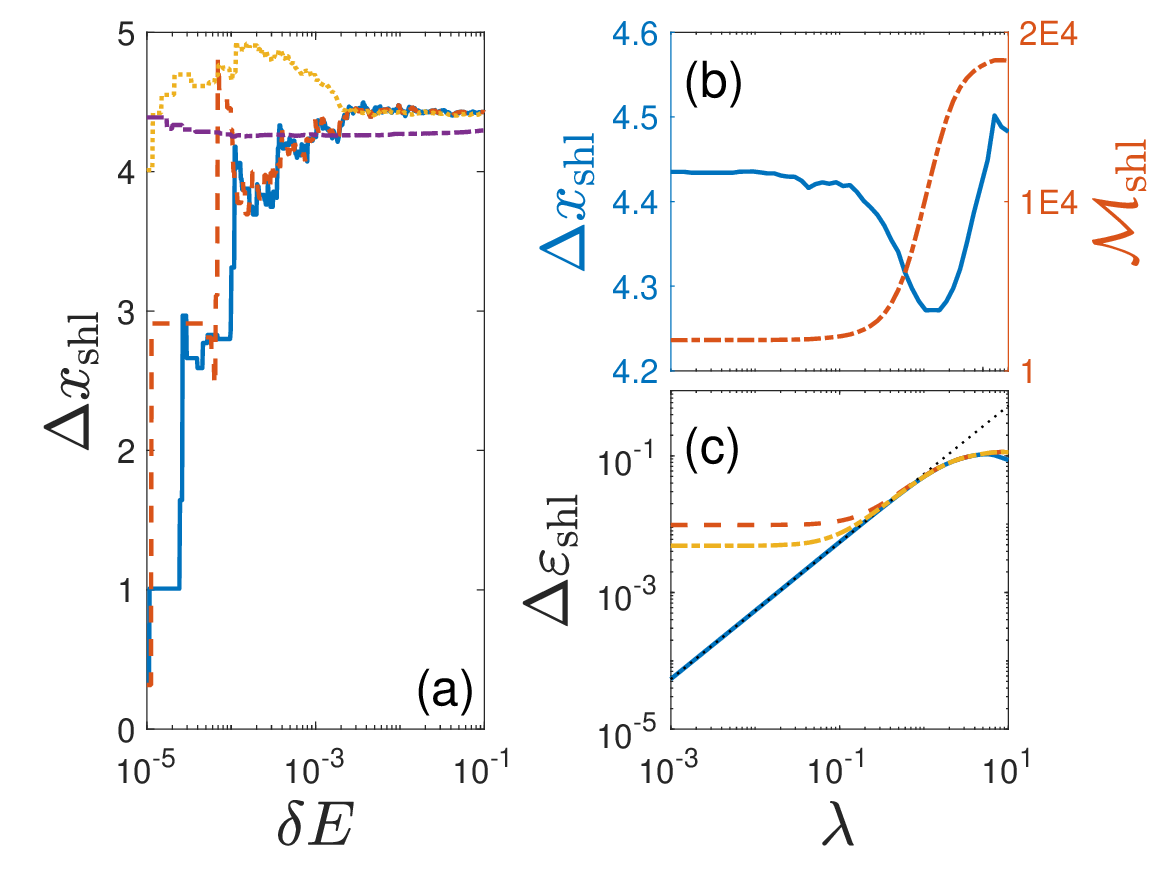}
\caption{
{\bf The width of the energy shell.} 
{\bf (a)}~The width $\Delta x_{\text{shl}}$ versus the width $\delta E$ of the arbitrary energy window about ${E=0.41666}$ for $\lambda=10^{-3}$ (solid blue line), $10^{-2}$ (dashed red), $10^{-1}$ (dotted orange), and $1.0$ (dash-dotted purple). The width of the energy shell is well defined (assuming quantum mechanically large but classically small $\delta E$) and the dependence of the obtained shell width on the coupling strength $\lambda$ is weak.
{\bf (b,c)}~The dependence of $\Delta x_{\rm shl}$, ${\cal M}_{\rm shl}$ and $\Delta \varepsilon_{\rm shl}$ on the coupling $\lambda$.  In the latter case one observe a residual value for small $\lambda$, corresponding to the arbitrary $\delta E$ (dashed line is for $\delta E=0.0333$ whereas dash-dotted line is for $\delta E=0.0166$). This numerical artifact can be eliminated by looking on ${\overline{\Delta \varepsilon_{\nu}}}$ (solid line), even if the eigenstates are non-ergodic! The dotted line is a guide to the eye, corresponding to linear $\Delta\varepsilon_{\rm shl}(\lambda)$ dependence
}  
\label{fg2}  
\end{figure}

\section{The energy shell}

Semiclassically, a microcanonical distribution occupies an energy shell of thickness $\delta E$ in phase-space. In the classical context the `energy surface' limit $\delta E\rightarrow 0$, is well defined. Quantum mechanically, the thickness $\delta E$ should be large with respect to the mean level-spacing but small with respect to the energy scale on which this spacing  varies, i.e. the shell width should be `classically small but quantum mechanically large'. 

It is practically useful to project the shell either over~$x$ or over $\varepsilon$ to obtain the shell probability distributions $p_x$ and  $p(\varepsilon)$. This means 
that $p_x \equiv \overline{p^{(\nu)}_x}$ and $ p(\varepsilon) \equiv \overline{p^{(\nu)}(\varepsilon)} $ are obtained by respectively averaging $p_x^{(\nu)}$ and $p^{(\nu)}(\varepsilon)$ over eigenstates $\nu$ that belong to the $\delta E$ energy window, i.e. ${|E_{\nu}-E| < \delta E/2}$. We define the dispersion $\Delta x_{\rm shl}$ as the second moment of the distribution $p_x$, as illustrated in \Fig{fg1}. Similarly, we extract the dispersion $\Delta \varepsilon_{\rm shl}$ from the unperturbed energy distribution $p(\varepsilon)$.  

In \Fig{fg2}, we plot the dependence of $\Delta x_{\rm shl}$ and $\Delta \varepsilon_{\rm shl}$ on the arbitrary chosen window $\delta E$ and the coupling strength $\lambda$. When $\delta E$ is large with respect to the level spacing, the shape of $p(x)$ becomes independent of $\delta E$, and only very weakly dependent on $\lambda$. Thus, the  dispersion $\Delta x_{\rm shl}$ is well defined.

By contrast, the energy dispersion $\Delta \varepsilon_{\rm shl}$ approaches the arbitrary window width $\delta E$ at small $\lambda$. This numerical artifact indicating that $\delta E$ is still too large. The dispersion $\Delta \varepsilon_{\rm shl}$ is the convolution of the individual widths $\Delta\varepsilon_\nu$ with the arbitrary $\delta E$. Physically meaningful result is obtained provided the contribution of $\delta E$ is negligible. It turns out that this numerical limitation can be avoided, because for classically small $\lambda$  we get 
\beq \label{eQCC}
\Delta \varepsilon_{\rm shl} = \overline{\Delta\varepsilon_\nu} 
= \lambda \sqrt{ [V^2]_{\varepsilon,\varepsilon} - [V]_{\varepsilon,\varepsilon}^2 }
\eeq
This holds even if the eigenstates are non-ergodic, as further discussed in the following subsections with consequences that are highlighted in \Sec{sFalse}. 
\Fig{fg2}c demonstrates the validity of this statement, and the implied linear dependence of $\Delta \varepsilon_{\rm shl}$ on~$\lambda$.

\subsection{Width of the energy shell}

Denoting unperturbed and exact eigenstates as $|\varepsilon\rangle$ and $|E\rangle$ respectively, the local density of states (LDOS) and the inverse-LDOS (the lineshape of the eigenstats) are defined respectively as follows: 
\beq
P^{(\varepsilon)}(E) &=& |\Braket{E}{\varepsilon}|^2
\\
p^{(E)}(\varepsilon) &=&|\Braket{\varepsilon}{E}|^2 
\eeq
What we call $\Delta \varepsilon_{\rm shl}$ 
is the $\varepsilon$-width of  $p^{(E)}(\varepsilon)$, 
while $\Delta_E$ is defined as the $E$-width of ${P^{(\varepsilon)}(E)}$.
Let us first clarify these definitions in a purely classical context. Assume that ${H(X,P)=H_0+\lambda V}$. 
Consider a microcanonical distribution ${H_0(X,P)=\varepsilon}$ and define  
${E = H(X,P)}$. We have the trivial identity 
\beq
\Delta_E^2 = \lambda^2 \left[ \braket{V(X,P)^2}_{\varepsilon}-\braket{V(X,P)}_{\varepsilon}^2  \right] 
\eeq
where the the $\varepsilon$ subscript indicates an $H_0$-microcanconical phase-space average. This result implies strict linear dependence  ${\Delta_E \propto \lambda}$. 
The dual statement for the inverse-LDOS is as follows.  
Consider a microcanonical distribution ${H(X,P)=E}$ 
and define ${\varepsilon = H_0(X,P)}$. 
We have the trivial identity 
\beq \label{eDeltaE}
\Delta \varepsilon^2 = \lambda^2 \left[ \braket{V(X,P)^2}_E-\braket{V(X,P)}_E^2  \right] 
\eeq
where the the $E$ subscript indicates an $H$-microcanconical phase-space average. Here there is an implicit $\lambda$-dependence of the latter. But if the difference between the surfaces ${H(X,P)=E}$ and the surfaces  ${H_0(X,P)=\varepsilon}$ is small, the $H$ microcanonical average can be calculated with an unperturbed surface of $H_0$, and the same result will be obtained, approximately, provided $\lambda$ is classically small enough.

\subsection{Second moment energy spreading}
\label{sQCC}

Here we provide the proof that the second-moment of the energy spreading has a robust linear dependence of $\lambda$, provided the perturbation is classically small, irrespective of quantum ergodicity. 
The proof is merely a translation of the classical statement \Eq{eDeltaE} into quantum language. 
We assume that ${H=H_0+\lambda V}$. Let $\ket{\varepsilon}$ be the eigenvalues of $H_0$, and let $\ket{E}$ be a chosen eigenstate of~$H$. Then it follows that    
\beq \label{eRqcc}
\Delta \varepsilon 
\ \ &=& \ \ \lambda \sqrt{ \BraKet{E}{V^2}{E} - \BraKet{E}{V}{E}^2 } 
\\
\ \ &\approx& \ \ \lambda \sqrt{ [V^2]_{\varepsilon,\varepsilon} - [V]_{\varepsilon,\varepsilon}^2 } \ \ \equiv \ \ \Delta_E
\eeq
with ${\varepsilon=E}$.
This result, which we further explain below, is independent of whether the eigenstate is quantum ergodic or not. The validity of this observation can be optionally verified by direct calculation of the second moment spreading for the non-ergodic perturbative eigenstates of \Eq{eFOPT}. 

Let us elaborate on the two steps in \Eq{eRqcc}. 
The first equality is based on a trivial identity 
that holds for the ${k=1,2}$ moments (and cannot be extended to higher moments due to operator ordering issue): 
\beq \nonumber
\braket{\varepsilon^k} 
&=& \sum_{\varepsilon}  |\Braket{\varepsilon}{E}|^2 \varepsilon^k  
\ \ = \ \ \BraKet{E}{H_0^k}{E} 
\\ \nonumber
&=& \BraKet{E}{(H-\lambda V)^k}{E} 
\\ \nonumber
&=& \left\{  
\amatrix{
E-\lambda \BraKet{E}{V}{E}  & |\text{for} \ k{=}1 \\ 
E^2-2\lambda E \BraKet{E}{V}{E} + \BraKet{E}{V^2}{E} & |\text{for} \ k{=}2  
}  \ \ \ \ \   
\right. 
\eeq
The second equality in \Eq{eRqcc} holds for a {\em classically small} $\lambda$, because the matrix elements of any generic operator between perturbed eigenstates is approximately equal to the matrix elements between the unperturbed eigenstates, namely 
\beq
\BraKet{E}{A}{E}  \ \approx \ \BraKet{\varepsilon}{A}{\varepsilon} \ \equiv \ A_{\varepsilon,\varepsilon} \ \ \text{with} \  \varepsilon \approx E
\eeq
This approximation is based on the observation that 
${\BraKet{\psi}{A}{\psi} = \trc[A \rho]}$  with ${\rho=\ket{\Psi}\bra{\Psi} }$ can be written as a phase-space integral over Wigner functions. If the microcanonical energy surface that supports $\rho$ is not related to the phase-space contours of $A$, the trace overlap is rather robust. A small perturbation to $\psi$ will not affect systematically the result (i.e. disregarding quantum fluctuations).

\section{Distinguishing ergodicity from chaos}

It is a common misconception to identify ``quantum chaos" with "quantum ergodicity". Quantum chaos is identified by the spectral statistics. In \Fig{fg3} we display a re-scaled  version of a common measure for level repulsion, namely the mean ratio of consecutive level spacings \cite{Atas13},
\beq
\tilde r\equiv \frac{\langle r\rangle-r_{\rm Poisson}}{r_{\rm GOE}-r_{\rm Poisson}}
\label{rmeasure}
\eeq
The value $\langle r\rangle=r_{\rm Poisson}=0.38629$ (${\tilde r}{=}0$) characterizes spectra that consist of uncorrelated levels. This applies to integrable systems but also to the uncoupled chaotic systems of our model.  
The value $\langle r\rangle=r_{\rm GOE}=0.53590$ (${\tilde r}{=}1$) holds for spectra of fully chaotic systems with Wigner-Dyson level spacing statistics.

The proper measure for the quantum ergodicity of the eigenstates, is the mean participation number, $\overline{\mathcal{M}_{\nu}}/\mathcal{M}_{\rm shl}$ (dashed line in \Fig{fg3}).
The $\mathcal{M}_{\nu}$ are calculated for the individual eigenstates, while $\mathcal{M}_{\rm shl}$ is calculated from the $p_{x,n}$ of the energy shell. For GOE statistics, a value of $\overline{\mathcal{M}_{\nu}}/\mathcal{M}_{\rm shl}\approx1/3$ indicates the ergodicity of the eigenstates. The $1/3$ factor emerges because the $p_{x,n}^{(\nu)}$ probabilities are not uniformly spread over the energy shell but statistically fluctuate.

Comparing the two measures, we note that the ergodic regime in \Fig{fg3} is much smaller than the chaotic regime indicated by spectral statistics. For a substantial $\lambda$ range the system is chaotic, yet its eigenstates are not ergodic. Similar discrepancies between chaos and ergodization measures have been previously observed without further exploration in \cite{Yurovsky11,Yurovsky23a,Yurovsky23b,Yurovsky25}

\section{The ETH ergodicity measures}

\begin{figure}

\includegraphics[width=\columnwidth]{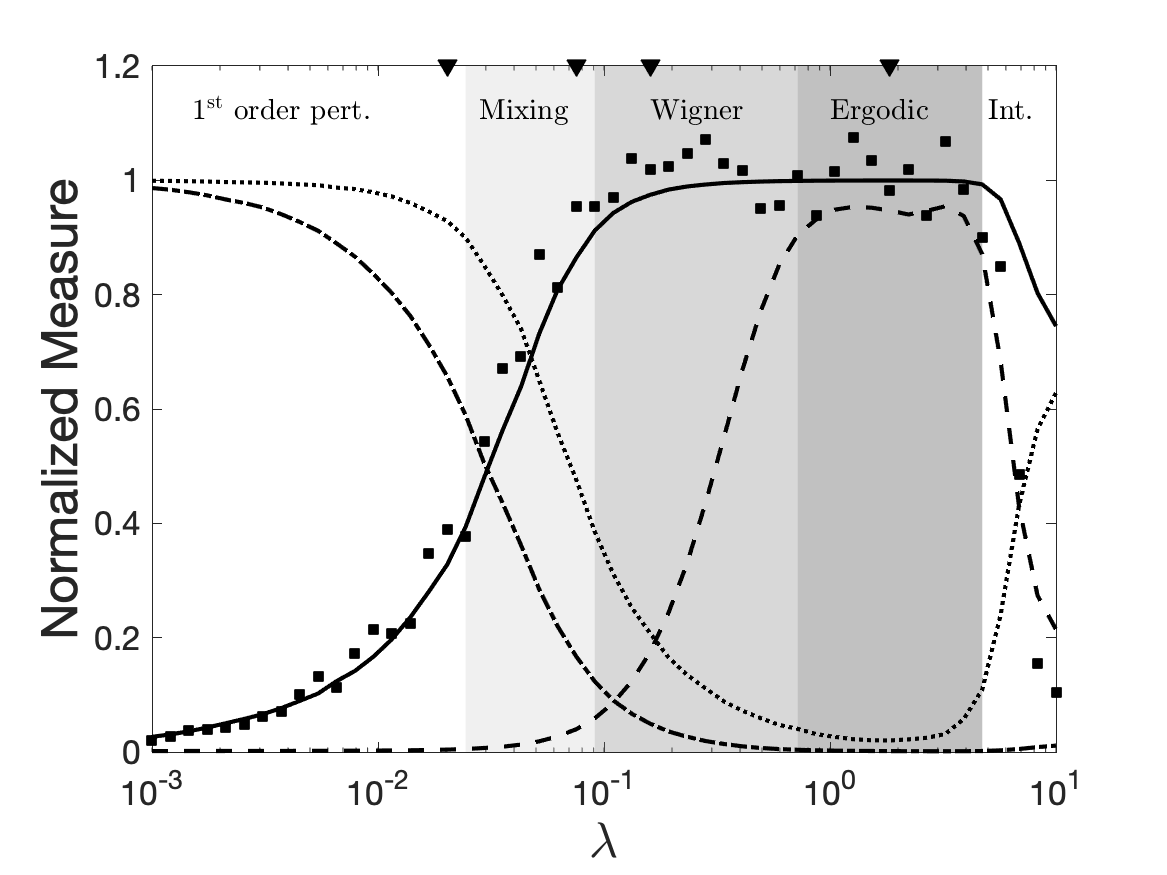}

\caption{{\bf Chaos and Ergodicity Measures.} 
Misc measures are calculated for the energy shell ${0.4<E<0.4333}$ as a function of the coupling strength $\lambda$, and normalized per the expected GOE values. 
The spectral measure for chaos {$\tilde{r}$} (square markers) is correlated with {$\overline{\Delta x_{\nu}}/\Delta x_{\rm shl}$} (solid line).  
The ETH measure {$\sigma_x /\Delta x_{\rm shl}$} (dotted line) is an insufficient measure for ergodization.   
The measure {$\overline{\mathcal{M}_{\nu}}/(\mathcal{M}_{\rm shl}/3)$} (dashed line) is an effective measure for the actual identification of the ergodic regime.   
We also plot the mean peak probability {$\overline{P_{\nu} }$} (dash-dotted line), which is used to identify the first-order perturbative regime (arbitrarily defined as $\overline{P_{\nu} }>0.6$).   
Inverted triangles mark the coupling strengths for the states presented \Fig{fg4}.
}  
\label{fg3}  
\end{figure}

The ETH predicts that for ergodic eigenstates $|E_{\nu}\rangle$ the expectation value of any local observable, such as $\braket{x}_{\nu}$, will exhibit low state-to-state fluctuations, with dispersion ${\sigma_x \propto 1/\sqrt{\mathcal{M}}}$. The mere decrease of state-to-state fluctuations however, is not a sufficient indication for ergodicity, as clearly demonstrated by comparing the earlier (weak coupling) drop of $\sigma_x$ with the later (stronger coupling) rise of $\overline{{\cal M}_\nu}$ in \Fig{fg3}.

\begin{figure}
\includegraphics[width=\columnwidth]{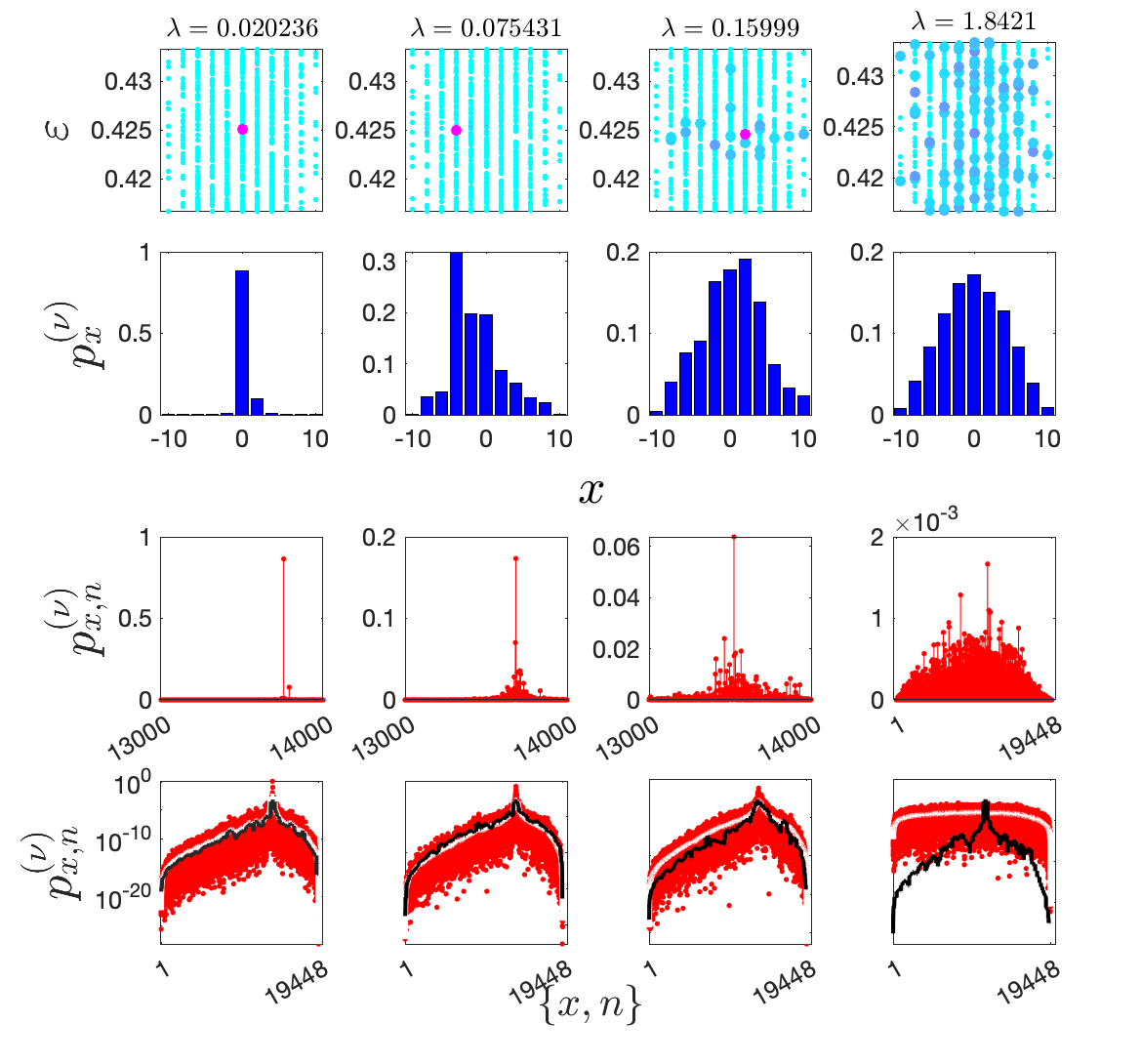}
\caption{
{\bf Representative eigenstates.} 
{\em Upper row:} Images of $p^{(\nu)}_{x,n}$ for a representative eigenstate with ${E\approx 0.425}$. Each panel is for a different coupling corresponding (from left to right) to the Perturbative, Mixing, Wigner, and Ergodic regimes. 
{\em Second row:} The resulting imbalance distribution $p_x^{(\nu)}$.
{\em Third row:} The $p_{x,n}^{(\nu)}$ probabilities are plotted versus $(x,n)$,  arranged in order of ascending unperturbed energy.
{\em Bottom row:} The same $p_{x,n}^{(\nu)}$ probabilities are compared on a log scale to the coarse-grained 1st-order perturbation profiles (solid lines). White lines are coarse-graining of $p_{x,n}^{(\nu)}$. 
}  
\label{fg4}  
\end{figure}

\section{Contrast with Wigner theory} 

Referring to the 2D energy landscape of the BHM, see \Fig{fg1}, the width $\Delta x_{\rm shl}$ is finite, with negligible dependence on $\lambda$, see \Fig{fg2}. This should be contrasted with the WBRM paradigm, where the horizonatal and the vertical dimensions are degenerate. Namely, $\Delta x_{\rm shl} = \Delta \varepsilon_{\rm shl}/\Delta_0 $, where $\Delta_0$ is the mean level spacing. Thus, in the WBRM model, $x$~can be viewed as a dimensionless energy coordinate. This observation leads to the understanding that the mathematical-oriented WBRM model is an oversimplification. The actual route towards thermalization in a realistic physical system like the BHM is more complicated and non-universal.

\section{The route to ergodicity}

We identify five distinct physical regimes in the intricate progression of the BHM towards ergodicity, as the coupling strength $\lambda$ increases. These are marked by the shaded regions in \Fig{fg3}, with a representative eigenstate for each in \Fig{fg4}. 
For the general discussion we note that the relevant model parameters are the mean level spacing $\Delta_0$, the bandwidth $\Delta_b$, and the typical value $\bar{V}$ of the in-band matrix elements. It is useful to define dimensionless bandwidth ${b=\Delta_E/\Delta_0}$, and dimensionless coupling  ${ \tilde{\lambda} := \lambda \bar{V}/ \Delta_0 }$.

\sect{The perturbative regime}
For very small values of $\lambda$, The peak probability $P_{\nu}$ is of order unity, and the $|E_{\nu}\rangle$ eigenstates can be described by 1st-order perturbation theory, which implies
\beq \label{eFOPT}
p_{x,n}^{(\nu)} = \left| \frac{ \BraKet{x,n}{\lambda V}{x_0,n_0} }{ \varepsilon_{x,n} - \varepsilon_{x_0,n_0} }\right|^2, \ \ \  (x,n) \neq (x_0,n_0) \ \ \ \ 
\eeq
where $(x_0,n_0)$ indicate the unperturbed eigenstate. This approximation holds as long as ${ \tilde{\lambda} <1}$. It has been argued in \cite{lds} that  \Eq{eFOPT} holds for the {\em tails} of the coarse-grained $p_{x,n}^{(\nu)}$ for larger values of $\lambda$, as long as quantum ergodicity is not yet attained. The validity of this claim for the BHM is demonstrated with striking accuracy in the lower panels of \Fig{fg4}. 

\sect{The Mixing regime} 
As $\lambda$ is increased nearby levels are mixed, leading to level repulsion reflected by the spectral measure~$r$. At this stage the eigenstates spread in the 'horizontal' $x$ direction of the ${(x,\varepsilon)}$ energy shell, yet remain localized in the 'vertical' $\varepsilon$ direction. A rough estimate for this horizontal mixing is provided by the standard Anderson model, predicting a localization length ${\xi \sim \lambda^2 }$. Full horizontal mixing is attained once ${\xi\sim N}$. This implies a border ${\tilde{\lambda}_{\rm mix} \sim \sqrt{N}}$ at which 'horizontal ergodicity' $\overline{\Delta x_\nu}/\Delta x_{\rm shl}\approx 1$ is attained.

\sect{The Wigner regime}
Once the horizontal mixing is achieved, we get into the Wigner regime of vertical spreading, where the WBRM theory applies.  A terse summary of the WBRM theory is provided in Appendix~A. 
The mixed levels form a core of vertical width ${\Gamma_E \propto \lambda^2 }$. This core determines the participation number $\mathcal{M}_{\nu}$.  
Full ergodicity is reached once ${\Gamma_E}$ becomes comparable with the vertical width of the energy shell \Eq{eQCC}. Consequently, the expected ergodicity threshold is ${\tilde{\lambda}_{\rm erg} \sim \sqrt{b}}$.

\sect{The Ergodic regime}
For $\lambda>\lambda_{\rm erg}$ there is full mixing within the energy shell, both horizontally and vertically.  This regime is semiclassical. To the extent it conforms to the WBRM model, we can say that in the Wigner regime ${ \Gamma_E \ll \Delta_E < \Delta_b }$, 
while in the ergodic regime ${ \Gamma_E \sim \Delta_E > \Delta_b }$, 
where $\Gamma_E$ indicates the width of the `core' region of non-perturbative mixing. 

\sect{The Integrable regime}
Finally, when ${\lambda \gg K}$ the intra-system hopping may be neglected and the system is reduced to the sum of disconnected integrable Bose-Hubbard dimers. Consequently, ergodicity is lost, $M_\nu$ and $\Delta\varepsilon_\nu$ decrease, and the state-to-state variance $\sigma_x$ increases as the ETH is violated. This regime is non-universal. In the WBRM the somewhat analogous regime is the Anderson regime ${\tilde{\lambda} > b^{3/2}}$, see Appendix~A for details.

\section{False ergodicity measures}
\label{sFalse}

Unlike $\overline{\mathcal{M}_{\nu}}/\mathcal{M}_{\rm shl}$, the ratio 
$\overline{\Delta \varepsilon_{\nu}}/\Delta \varepsilon_{\rm shl}$ 
can not be used as an ergodicity measure. In fact, the first equality in \Eq{eQCC} implies that the latter ratio is of order unity, irrespective of ergodicity. As discussed in \Sec{sQCC}, a straightforward argument shows that ${\Delta \varepsilon_{\nu}}$ for any eigenstate is determined by the strength of the perturbation $V$, and is linear in $\lambda$. This is true even if the eigenstate is localized within the energy shell. The most dramatic demonstration is based on \Eq{eFOPT} for which ${ \mathcal{M} \sim 1 }$, but nevertheless the dispersion ${\Delta \varepsilon_{\nu}}$  is determined by the tails, and is not affected by ${ \mathcal{M} }$. Consequently,  the linear $\lambda$ dependence of $\overline{\Delta \varepsilon_{\nu}}$  is {\em insensitive} to both the ${\tilde{\lambda}_{\rm mix} }$ and the ${\tilde{\lambda}_{\rm erg} }$ thresholds, hence it cannot be used as an ergodicity measure.

\section{The regime borders} 

Consider the scaling of the mixing- and ergodicity borders $\lambda_{\rm mix}$ and the $\lambda_{\rm erg}$, with~$N$. Within a semiclassical treatment a rough estimate for the localization length is ${ \xi = (1/\Delta_0)D_x }$, where the diffusion coefficient in $x$ is ${ D_x = N^2 D_{cl}(\lambda) }$. 
Note that $D_{cl} \propto [\lambda V_{cl}]^2$, where $V_{cl}$ is the strength of perturbation in `classical' units. The density of states is ${(1/\Delta_0) = N^d \Omega_{cl} }$, where the number of degrees of freedoms  is $d=M{-}1$, and $\Omega_{cl}$ is the phase-space area of the energy shell.  
We note that $\xi$ is possibly larger as discussed in \cite{mlc}.   
From ${\xi < N}$ we get the condition  
\beq 
\Omega_{cl} D_{cl}(\lambda) < (1/N)^{1+d}  
\ \ \ \ \leadsto \ \ \ \ \lambda_{\rm mix}
\eeq
Turning to the ergodic border, the condition is ${\Gamma_E < \Delta_b}$.
The Feingold-Peres relation \cite{mario1,mario2,mario3,mario4} (equivalent to ETH) is Eq.6 of \cite{lds} that relates matrix elements to fluctuations. It is implied that $\Delta_b=\hbar/\tau_{cl}$, and that ${\Gamma_E = (1/\Delta_0) \lambda^2 \bar{V}^2 = \lambda^2 V_{cl}^2 \tau_{cl}/\hbar }$. Here $\hbar=1/N$. 
Thus we get the condition 
\beq 
\lambda^2 V_{cl}^2 \tau_{cl}^2 < (1/N)^2 
\ \ \ \ \leadsto \ \ \ \ \lambda_{\rm erg}
\eeq
Comparing the two conditions we conclude that for large $N$ the mixing border is typically encountered before the ergodic border.

\section{Summary}

There are numerous studies where thermalization of one-dimensional spin-chains is considered  \cite{Chain1,Chain2,Chain3,Chain4}. The perturbation $\lambda V$ in such models is responsible for non-integrability, and the length $L$ comes into consideration within a framework of a finite-size scaling scheme. The paradigm that we use for thermalization is quite different. It is motivated by the realization that thermalization of large disordered system can be modeled as the ergodization of small nearby chaotic subsystems, that exchange particles and energy, sometimes referred to as hot spots \cite{Basko}. 

No special assumptions about the subsystem geometry are made in our model, and $L$ is not a physically meaningful geometrical parameter for the analysis. The number of sites $M_{\text{subsystem}}$ should be larger than~2 in order to have chaos, or better larger than~3 in order to avoid Kolmogorov-Arnold-Moser phase-space barriers (we used ${M_A=M_B=4}$). The spectral structure illustrated in \Fig{fg1}(b) as well as the distinction between horizontal and vertical ergodization demonstrated in \Fig{fg4} are not system-specific and apply to all bipartite composite systems, regardless of their geometry or size. This point is further demonstrated in Appendix~B. What really counts is the Hilbert space dimension, and hence the number of particles ($N$) has appeared in our discussion.  

RMT modeling provides a partial understanding of the {\em route} towards ergodization. For physical subsytems a more careful definition of the energy-shell concept is required. Then it becomes clear that two distinct steps are involved, with different thresholds $\lambda_{\rm mix}$ and $\lambda_{\rm erg}$, that signify spectral quantum chaos and ETH ergodization respectively.  

\ \\
{\bf Acknowledgments} --  
AV acknowledges support from the NSF through a
grant for ITAMP at Harvard University. DC acknowledges support by the Israel Science Foundation, grant No.518/22.

\appendix

\section{The WBRM model}

The WBRM can be written schematically as follows:
\beq
H = \text{diag}\{ \varepsilon_n \} + \text{offdiag}\{ W_{n,m} \}
\eeq
Parameters: 
\beq
\Delta_0 &=& \text{level spacing}  \\
\Delta_b  &=& \text{range of hopping} \\ 
\bar{W} &=& \text{strength of hopping}
\eeq
Dimensionless parameters:
\beq
b &=& \frac{\Delta_b}{\Delta_0} 
\ = \ \text{dimensionless band width} 
\\
\lambda &=& \frac{\bar{W}}{\Delta_0} 
\ = \ \text{dimensionless coupling} 
\eeq
The Wigner regime is 
\beq
1 < \lambda < \sqrt{b} 
\eeq
In this regime there are two ``width" scales. One is $\Gamma_E$ defined via PN calculation, and the other is $\Delta_E$
defined via a second moment calculation. One obtains   
\beq
\Gamma_E &=& 2\pi \frac{\bar{W}^2}{\Delta_0} 
\\ 
\Delta_E &=& \sqrt{b} \bar{W} 
\eeq
Accordingly, 
\beq
\frac{\Gamma_E}{\Delta_0} &=& 2\pi \lambda^2 
\ = \ \text{dimensionless core width} 
\\
\frac{\Delta_E}{\Delta_0} &=& \sqrt{b} \lambda 
\ = \ \text{dimensionless shell width} 
\eeq
In the non-ergodic Wigner regime 
${ \Gamma_E \ll \Delta_E < \Delta_b }$, 
while in the ergodic regime ${ \Gamma_E \sim \Delta_E > \Delta_b  }$.
Note that in the RMT model there is also an Anderson localization regime ${\lambda > b^{3/2} }$, where the $\Gamma_E/\Delta_0$ width saturates to the value ${\Delta_{\xi}/\Delta_0 = \xi \sim b^2}$, that is much smaller than the unbounded semiclassically expected result ${\Delta_E/\Delta_0 \propto \lambda}$.   

\begin{figure}[t!]
\includegraphics[width=\columnwidth]{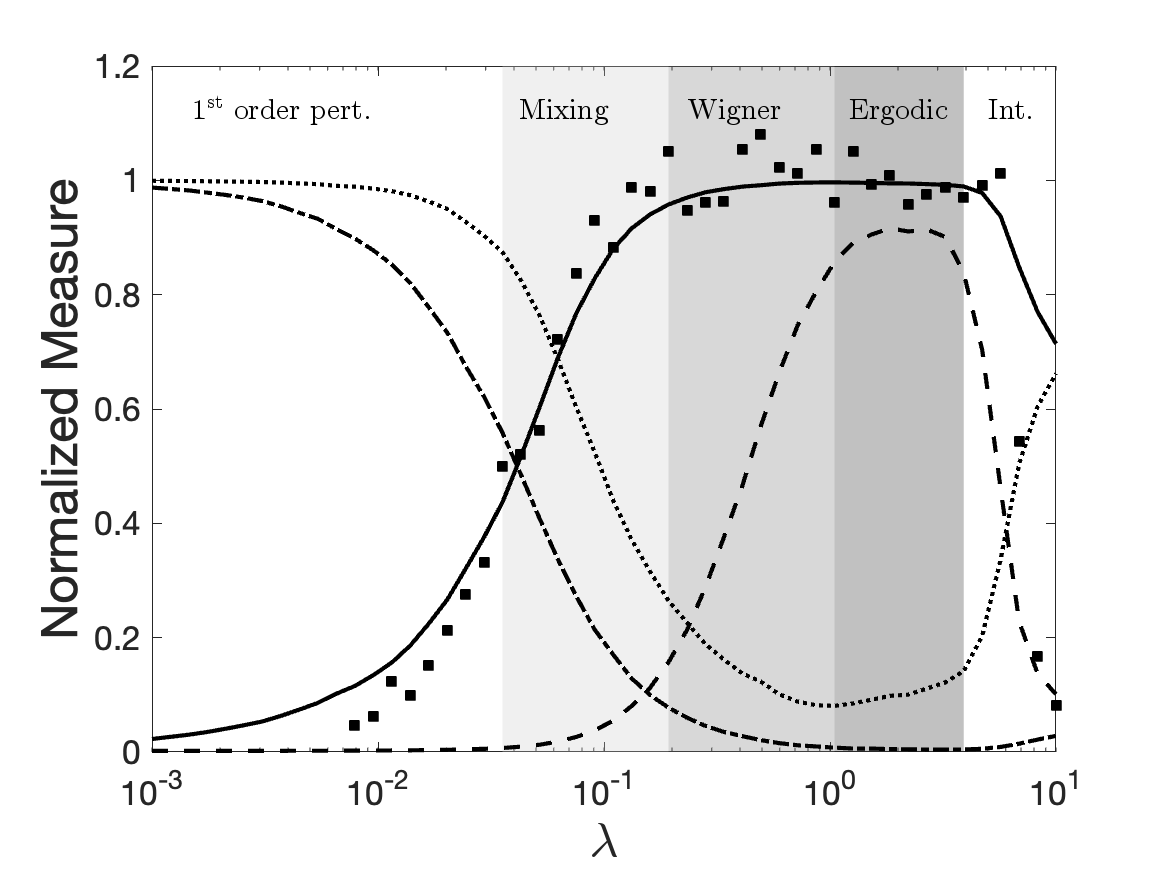}
\caption{{\bf Chaos and Ergodicity Measures.} 
Same as \Fig{fg3} but with random internal couplings between all the sites of the coupled subsystems, selected from a normal distribution with zero mean and unit standard deviation.  Thus, each subsystem is not a chain, but a fully connected cluster of arbitrary geometry. 
}  
\label{fg3sm}  
\end{figure}

\section{Irrelevance of geometry}

In order to demonstrate the irrelevance of geometry in the context of our analysis, we have generated numerical results for coupled subsystems, where each subsystem is a fully connected cluster rather than a chain.  
The results are presented in \Fig{fg3sm}. All the main features of \Fig{fg3} and, in particular, all five coupling regimes remain unchanged.

\ \\


\end{document}